# Characterizing solute hydrogen and hydrides in pure and alloyed titanium at the atomic scale

Yanhong Chang[1], Andrew J. Breen[1], Zahra Tarzimoghadam[1], Philipp Kürnsteiner[1], Hazel Gardner[2], Abigail Ackerman[3], Anna Radecka[4], Paul A. J. Bagot[2], Wenjun Lu[1], Tong Li[1,5], Eric A. Jägle[1], Michael Herbig[1], Leigh T. Stephenson[1], Michael P. Moody[2], David Rugg[4], David Dye[3], Dirk Ponge[1], Dierk Raabe[1], Baptiste Gault[1*]

[1] Max-Planck-Institut für Eisenforschung GmbH, Max-Planck-Straße 1, 40237, Düsseldorf, Germany.

[2] Department of Materials, University of Oxford, Parks Road, Oxford, OX1 3PH, United Kingdom.

[3] Department of Materials, Royal School of Mines, Imperial College, Prince Consort Road, London SW7 2BP, United Kingdom.

[4] Rolls Royce plc. PO Box 31, Derby, United Kingdom.

[5] Institut für Werkstoffe, Fakultät für Maschinenbau & ZGH, Ruhr Universität Bochum, Germany.

## Abstract

Ti has a high affinity for hydrogen and are typical hydride formers. Ti-hydride are brittle phases which probably cause premature failure of Ti-alloys. Here, we used atom probe tomography and electron microscopy to investigate the hydrogen distribution in a set of specimens of commercially pure Ti, model and commercial Ti-alloys. Although likely partly introduced during specimen preparation with the focused-ion beam, we show formation of Ti-hydrides along α grain boundaries and α/β phase boundaries in commercial pure Ti and α+β binary model alloys. No hydrides are observed in the α phase in alloys with Al addition or quenched-in Mo supersaturation.

## 1   Introduction

Ti-alloys exhibit excellent ratios of weight to fatigue-allowable-strength, high thermal stability, toughness, biocompatibility, corrosion resistance [1], and can form stable hydrides with moderate formation enthalpies and has potential for solid-state hydrogen storage [2]. It has long been observed that titanium alloys easily adsorb large amounts of oxygen and hydrogen [3], and Ti alloys usually undergo a hydrogen removal heat treatment during processing in order to lower the residual H content. This step supposedly alleviates concerns related to H in solid solution or as hydride to facilitate crack initiation or propagation which would be limiting the life-time of Ti alloys components routinely used in safety-critical products such as biomedical implants, jet engines and airframes, satellites, space vehicles, chemical and desalination plants. Pure Ti has two allotropic forms: a body-centered cubic β structure above 882°C, and, below, a hexagonal close-packed α structure. Solutes such as Al, V, Sn, Zr, Mo, Fe, as well as O and H have different affinities for α and β, modifying the properties. Ti alloys are interesting from the standpoint of alloy design because the proportions of the two phases can be adjusted at different stages in processing, allowing the morphological distribution of the phases to be adjusted to tailor the properties for different applications. Hence, the presence and partitioning of H in the bulk of a phase, at or near crystalline defects, within the microstructure of Ti-alloys are of both fundamental interest and profound social relevance, yet direct observations of hydrogen at the nanoscale remain elusive but are crucial to understand how H can so strongly affect materials' properties.

* Correspondence and requests for materials should be addressed to BG (b.gault@mpie.de)

The role of H is usually indirectly inferred from changes in materials' properties [4] and not from direct observations, making unravelling mechanisms particularly challenging. In metals, H is either in solution, trapped at crystal defects or in the form of hydrides, each with a distinct influence on ductility, fracture, softening, hardening, and creep [5–7]. Many mechanisms [8–11], collectively referred to as *hydrogen embrittlement,* have been proposed, but how H *within* materials interacts with microstructural features e.g. crystal defects, interfaces or secondary phases remains elusive [4]. In Ti-alloys, H partitions and stabilizes the β-phase and is sometimes a temporary alloying element that enhances processability [12,13]. However, H also causes embrittlement and Ti-alloys usually undergo a dedicated heat treatment to lower residual H and alleviates concerns pertaining to crack initiation [14] or propagation that limit component's lifetime.

Precise characterization of the H distribution in engineering alloys is extremely challenging. Electron microscopy techniques, in particular transmission electron microscopy (TEM) has been deployed to reveal how H affects defects' behaviour in metals [4] or to image atomic columns in a stable hydride [15]. However, because H lacks an electron outer shell, it can only be observed via plasmon interactions in Electron Energy Loss Spectroscopy (EELS) and so TEM is yet to successfully image solute H within an industrial alloy. Atom probe tomography (APT) exploits the effect of an intense electric field applied to a sharp, needle-shaped specimen to trigger the emission and analyse individual atoms, providing three-dimensional analytical mapping of a material with sub-nanometre resolution[16]. APT is the only technique which can currently locate, with sub-nanometer spatial resolution, and quantify hydrogen within a microstructure. The difficulty lies in that APT datasets typically register 0.5 – 1.5 at% [17] of spurious H originating from residual gases in the ultra-high vacuum chamber. Deuterium charging, as a surrogate for H, has shown promising results recently [18–21], but even so, the presence of residual $H_2$ in the mass spectra will affect the results and this has precluded routine quantitative assessment of hydrogen in metals by APT. Here, we show how APT, also combined with electron microscopy, still provides insights into the partitioning of hydrogen and the formation of hydrides in binary model and engineering Ti-alloys processed through conventional metallurgical routes and additive manufacturing.

## 2 Materials and methods

Commercially-pure Ti used here was of Grade 2. The two binary alloys Ti-4wt.% Mo and Ti-2wt.% Fe were synthesized in an electric arc furnace under Ar atmosphere and cast as a rectangle billet in-house at MPIE. As-cast Ti-2Fe billet was heated in the β region at 950°C for 13h under vacuum ($5*10^{-2}$ Pa) and then quenched by helium in 5 seconds via dilatometer. The as-cast Ti-4Mo was homogenized at 950°C for 24 hours followed by quenching, and then annealed at 600°C for 2h for β reversion.

The Ti64 alloy was processed by Selective Laser Melting using a SLM Solutions 280HL machine under an Ar atmosphere at an energy density of 145 J/mm3 from gas atomized, grade 23 Ti64 powder. The sample was investigated in the as-produced state. According to the manufacturers specifications grade 23 may contain max. 0.0125wt% H. The Ti6246 alloy is β rolled, then α-β forged with controlled cooling using a gas fan quench. The sample was then heated at 960°C and cooled at 7°C·min$^{-1}$ to 800°C, followed by water quenching to dissolve secondary and retain primary α.

The bulk hydrogen content in these four materials was therefore also measured in a Leco Inc. RH 402-Hydrogen Determinator coupled with a HF 402 induction furnace. The results are 0.175 at.%, 0.153 at.%, 0.151 at.%, 0.0319 at.% respectively for the as-cast Ti4Mo and Ti2Fe, the Ti64, and the Ti2Fe after heat-treatment and quench.

Scanning electron and electron back-scattered diffraction (EBSD) micrographs were acquired on a JEOL JSM at 15kV acceleration voltage. Specimen for APT were prepared from the surface of mechanically polished Ti samples by in-situ lift-out in a scanning-electron microscope / focused ion beam, following the protocol outlined in ref. [22] including the final steps at low acceleration voltage (14kV) to limit the damage caused by high-energy ions. In addition, upon finalising the needle-shape, a final cleaning of the specimen at 2kV acceleration voltage was performed so as to remove the region of the material likely affected by implantation damage. Two different microscopes were used: FEI Helios Nanolab 600 and 600i.

The CP-Ti, two binary alloys, Ti-Mo and Ti-Fe, and Ti6246 were analysed on a Cameca LEAP 3000 HR, operated in high-voltage pulsing mode with 15% pulse fraction, 200kHz pulse frequency and 0.3% target evaporation rate at a base temperature of 60K. The pressure in the ultrahigh vacuum chamber was consistently below $6\times10^{-11}$ mbar. Finally, the Ti64 on a Cameca LEAP 5000 in HV pulsing mode with a pulse fraction of 15% at a repetition rate of 200kHz, at a base temperature of 80K ± at 1.5 ions per 1000 pulses, and a pressure in the ultrahigh vacuum chamber below $8\times10^{-11}$ mbar. High-voltage pulsing mode was chosen to ensure that data from multiple specimens and material systems could be compared.

TEM investigations, prior to APT, were conducted on APT specimens in a Phillips CM20 with LaB$_6$ filament operated at 200kV. The experimental setup was described in ref [23].

# 3 Results
## 3.1 Ti-4Mo

Figure 1(a) and Figure 1(b) show the microstructure of the as-quenched Ti-4Mo alloy, a mixture of massive and acicular α' martensite with an hexagonal structure, consistent with previous reports [24,25]. Inside the massive martensite colonies, almost parallel α' laths are observed which belong to the same variant. The H and Mo elemental distribution revealed by the tomographic reconstruction obtained from an APT analysis that contains three grains and two boundaries is displayed in Figure 1(c). The detector maps extracted from the three different grains each reveal a pattern with a 6-fold symmetry corresponding to the base-plane of the hexagonal lattice as shown Figure 1(c). The position of the crystallographic poles revealed by these maps allows to derive the crystallographic orientation of each of the three grains. The change in the position of the basal poles indicates the change in orientation between grains [26]. In between the top grain and the middle grain, the basal pole almost overlap, which means they are the same variant with almost similar orientation, and indicates that the top grain boundary is likely a lath boundary. On the contrary, the shift of the pole position from the middle grain to the bottom grain is quite large, which indicates misorientation between the middle grain and the bottom grain is rather large. The partial crystallographic information available in such dataset is insufficient to perform a full quantitative assessment of the misorientation, but based on the distance between the main basal pole and the typical field-of-view of the instrument, the angle between the two (0002) directions is between 28 and 32 degrees. This makes the bottom grain boundary is either of a different variant within the same prior β-grain or from another prior β grain. The corresponding composition profile shows a slight Mo segregation along both boundaries, and a nearly homogeneous Mo distribution in the α' martensite. The Mo segregation along the martensite lath boundary indicates the segregation occurred after martensitic transformation during the quenching process. The amount of Mo segregation at these boundaries appears not directly related to the misorientation angle of the boundaries. Previous atom probe and electron microscopy work demonstrated the complexity of the relationship between interface misorientation and the associated segregation content (see e.g. [27]). In addition, the martensitic transformation sequence likely plays an important role.

The Ms temperature of Ti-4Mo alloy is very high, namely, 973K [25]. The observed mixture of massive and acicular martensite in the current microstructure supports this fact and implies that the water-quenching is insufficiently fast to prevent diffusion during the transformation. Over the temperature range (Ms - Mf) of the martensitic transformation, the amount of segregation may thus rather relate to the martensite transformation sequence instead to a Gibbsian effect related to a single misorientation value. Larger amounts of segregation may thus occur at boundaries of those martensite laths that transformed at a higher temperature than those that transformed at lower temperature [28]. In addition, the martensitic transformation mechanism may cause asymmetric lath structures with unequal boundaries and different dislocation content on adjacent borders of the same martensite lath [29]. This may also contribute to the different amount of Mo segregation on the two boundaries shown in Figure 1, as well as to the different profiles of segregation observed on either side of the boundary.

The average H composition in the data is approximately 2 at.%, with a slight enrichment up to 3.5 at.% at the lath boundary. After 15min at 600°C, the alloy is comprised of reverted metastable β and martensite. Figure 2 shows that Mo preferentially enriches up to 5-7at.% Mo at a boundary, which probably is a thin layer of metastable β, surrounded by 5-10 nm Mo-depleted (≤ 1at.%) regions on both sides due to limited diffusion at 600°C for 15 min. Inside the α' lath, the Mo concentration is around 2 at.% as in the as-quenched condition. APT reveals that the β layer contains up to 22 at.% H, while only 1.5-2 at.% H is measured in the surrounding α' martensite, consistent with H levels in the as-quenched case.

Following further β-reversion of Ti-4Mo after 2h at 600°C, the supersaturated α' martensite is fully decomposed into an α+β phase mixture. Figure 3(a) is a bright-field micrograph and Figure 3(b) the corresponding electron diffraction pattern of an APT tip containing a hydride formed at an α/β interface. The diffraction patterns are taken from the $[\bar{1}11]\beta$ // [011]hydride zone axis, as confirmed by simulated patterns in Figure 3(c). Figure 3(d) is a colored dark-field micrograph, generated from the $(\bar{1}1\bar{1})$ hydride and $(01\bar{1})\beta$ spots circled in Figure 3(b), that allows for direct correlation with the reconstructed APT map displayed in Figure 3(e). The Ti phases are easily differentiated by their Mo concentration of approx.13 at.% Mo in β phase and less than 0.1 at.% in the α phase (cyan). 4 at.% H was detected in β while < 1at.% H in α, which is consistent with residual levels reported in the literature [17]. The hydride between α and β phase contains 50±2 at.% H, as revealed from the one-dimensional concentration profile in Figure 3(f). This composition agrees with previous reports by Takahashi et al. [30]. Correlative TEM/APT reveals that the hydride was already present prior to APT measurement, rather than formed as the specimen was cooled down for APT analysis, or induced in the APT chamber.

### 3.2 CP-Ti and Ti-2Fe

To confirm that these observations were not related to a single alloy system, another alloy system Ti-2Fe was investigated, and in a second step, in single phase α-Ti grade 2 commercially-pure titanium (CP-Ti). First, investigations were performed on a two phase Ti-2wt.% Fe alloy synthesized following the same protocol as the reverted Ti-4 wt.% Mo. A hydride formed at an α–α grain boundary, shown in Figure 4, was analysed by atom probe crystallography [31]. The detector histograms in (a), obtained during the analysis, reveal varying pole figures which evidence a change in crystal structure and orientation. The poles of both top and bottom grains exhibit a 6-fold symmetry characteristic of the basal planes in the hexagonal α phase. The shift in the position of this pole between the top and the bottom α grain is equivalent to a change in orientation of 5.4° between the two grains. The 3-fold symmetry of the pole in the middle phase that contains approx. 50 at.% H, identifies the {111} pole of the hydride. The {111} pole of the hydride and the {0001} pole of the bottom α grain appear superimposed, as expected from the pattern in the maps in Figure 3a. This indicates that the interface between the hydride and the α grain is semi-coherent: {0001}α // {111}hydride, which is consistent with the previously reported orientation relationship between α-Ti and the γ-hydride [32].

Second, specimens of CP-Ti were prepared under the same conditions. The APT measurement were conducted with the same experimental parameters. The presence of hydride was observed in almost all the specimens, one of which is shown in Fig. 5a. Crystallographic analysis based on the pole figures indicates that the hydride has the same orientation relationship with α-matrix as shown in Fig. 4. A TEM lamella of CP-Ti was also prepared under the same conditions and hydrides were observed in the form of thin platelets, as seen in the bright-field micrograph in Fig. 5b. The diffraction pattern along $[110]_{hydride}$ // $[01\bar{1}0]_{\alpha}$ demonstrates the crystallographic orientation relationship between the hydrides and α-Ti being $\{0002\}_\alpha$ // $\{002\}_{hydride}$. The superlattice reflections from $(0\bar{1}1)$ indicates that the hydride is γ-hydride with a face-centered tetragonal lattice (fct) and an ordered arrangement of hydrogen, consistent with the hydride reported in reference [32] and [33]. Fig.5c is the corresponding dark-field image generated from the $(11\bar{1})\gamma$ spot marked in yellow in the diffraction pattern in the inset in Fig. 5(b) that shows only the hydrides within the α-Ti.

### 3.3 Ti6246, Ti64

To further investigate the effect of alloying in the α phase on the hydrogen solubility and hydride formation, two commercial alloys, namely Ti6246 and Ti64, both with Al addition, an α stabilizer, were analysed by APT as shown in Figure 6(a–b). Both materials are two phase alloys, with the α phase solute solution strengthened by 10-12 at. % Al. Figure 6(a) shows an α-Ti plate precipitated from β-Ti during heat treatment of Ti6246, with approx. 3 at.% H in α and 17 at.% in β. As shown in Figure 6(b), in Ti64, 2–3 at. % H enrichment on α lath boundaries and up to 13 at. % inside β blocks are detected along with V segregation and partitioning. The α alloyed with Al contains 1.5–3 at.% H, similar to the Mo-supersaturated martensite (Figure 1) and slightly higher than in unalloyed α (Figure 5). Neither of these alloys exhibited any hydride.

## 4 Discussion
### 4.1 Hydrogen introduction

The bulk hydrogen content in the bulk materials was consistently 10–30 times lower than those measured by APT. H was likely introduced during the preparation of specimens. Absorbed water or acid during mechanical or chemical polishing can cause absorption of H and the formation of hydrides [34]. Ding's work clearly showed that more hydride formed in the FIB-prepared TEM foil cut from the water/acid solution polished bulk sample than from the a dry-ground sample [29]. Adsorbed water or acid during mechanical or chemical polishing can facilitate hydrogen penetration within the sample.

Specimens for this study were prepared exclusively by FIB at room temperature, at approx. $10^{-6}$ mbar. The residual gas of the FIB vacuum chamber, e.g. hydrocarbons and/or moisture, probably is a major H source. Carpenter et al [35] pointed out that energetic ion beam (5keV broad Argon ion beam) might stimulate the decomposition reaction of hydrocarbon and moisture at metal surfaces and hence produce more hydrogen. Moreover, Wnuk et al [36] revealed that 500 eV electron beam irradiation could induce the decomposition of the organometallic precursor for Pt deposition, which is accompanied by the evaporation of methane and hydrogen gas phase. Therefore, it is reasonable to assume that another major source for H uptake during FIB preparation is the hydrogen gases generated by electron/ion beam irradiation of the organometallic gas precursor.

Hydrogen uptake would be aided by the high affinity of Ti for hydrogen and the rapid in-diffusion rate of hydrogen enhanced by the high reactivity of the sputtered surface and beam-induced heating. Ding and Jones suggested that FIB-induced point defects trigger or accelerate the formation of hydrides during the preparation of TEM specimens [33]. In comparison to TEM, APT specimens have a lower surface to volume ratio, smaller amount of Pt deposition and are prepared at lower acceleration voltage, current and dose, taking the final cleaning step of APT specimens at 2-5 kV into account. It is still likely that significant amounts of H were introduced during the annular milling steps of the specimen preparation protocol. This explains that averaging over eight APT datasets obtained for the Ti-2Fe alloy results in approximately 5 at.% H, well above expected levels.

In the future, strategies could be pursued to preclude H uptake during sample preparation and transportation, for instance, using water- and acid-free methods for the polishing of bulk samples; using cryo-FIB to limit H in diffusion during preparation; improving the level and quality of the vacuum in the FIB and vacuum transport specimens between instruments; provide a gentle heat treatment of the specimens post preparation to remove absorbed H as well as possible hydrides formed. This latter point will not rid the specimen of the possible microstructure evolution from treatments at high temperature, but it would also not necessary heal the specimen from modifications caused by the formation of the hydrides, e.g volume expansion and dislocation generation. Although a significant proportion of the H likely comes from the preparation of the specimens, our observations still provide unique insights.

### 4.2 Hydrogen quantification

Strong variations in the quantity of hydrogen detected within an atom probe experiment can usually be predominantly correlated with variations in the residual pressure within the ultra-high-vacuum chamber of the microscope [37], but also possible differences in the local composition and structure. The chamber pressure is monitored during the analysis, and only minute changes are observed from dataset to dataset, and the base pressure during the measurement is consistently in the range of $4 - 6 \times 10^{-11}$ mbar. Admittedly, the partial pressure of hydrogen is not individually measured, it is unlikely that it would significantly fluctuate without being accompanied with an increase in the overall pressure.

Regarding the other two aspects, changes in the amount of hydrogen detected can also usually be linked to a change in the local strength of the electrostatic field, which in turn can be made apparent by tracking the relative ratio between the different charge state of e.g. Ti (i.e. $Ti^{3+}/Ti^{2+}$) as extensively discussed by Kingham and co-workers [38,39]. A higher concentration of hydrogen from the residual gas is usually found when the electrostatic field is low, as shown in the work of Tsong – see for instance ref. [37].

In the case of the alpha/beta microstructure, the level of H in alpha is often in the range that is expected from the background level 1–2 at%, whereas in beta, it is consistently higher, up to 10–15 at%. This is evidenced in Figure 7, where the relative amounts of $H^+$, $H^{2+}$ are plotted across an α/β interface in the dataset presented in Fig 6(a) for the Ti6246 alloy. The variations in the Ti charge-state ratio is also plotted (in blue). The electric field is clearly higher in beta, with an increase in the absolute amount of $Ti^{3+}$ as well as in the ratio between $Ti^{3+}$ and $Ti^{2+}$. Based on the reports by Kellogg [40], a higher field should result in a lower amount of H originating from the residual gas. Therefore, the detected background H in beta should be lower than in alpha, and with values measured of up to 10-15 at.%, the H in beta must be mostly from solute hydrogen.

### 4.3 Hydrogen distribution

In a sense, the damage and implantation of H induced by the FIB accelerates the maturation process of the specimen's microstructure, which, during its lifetime in service, may progressively pick up H. These observations provide insights on the distribution of H within materials, i.e. partitioning between phases and at interfaces. As summarised in Figure 8, for Ti-2Fe and Ti-4Mo, ≤1 at.% H is usually detected in α-Ti that has a low H solubility [41]. Such levels are consistent with residual gas in APT. This contrasts with 5–7 at.% H in equilibrated β after full annealing. We reveal hydride formation at α/β interfaces and α lath boundaries, which agrees with the model proposed for hydride film growth at interfaces that provide rapid transport of H, enhance segregation and thus accelerate hydride nucleation and growth [42]. Metastable β contains 10–25 at.% H, higher than in equilibrated β, likely linked to large amounts of defects. Interestingly, Figure 7 shows higher H solubility in the alloyed α-phase but no hydrides even at α/β phase boundaries and α grain boundaries, even though up to nearly 20 at.% H is dissolved in the adjacent β.

### 4.4 Alloying elements suppressing hydride formation

Hydride formation causes volume expansion, which must be accommodated by both elastic and plastic deformation. Therefore, the strength of the matrix is a prominent parameter that hinders hydride formation [41].The α' martensite presented in Figure 1(c) is solid solution strengthened by Mo, which probably requires greater supersaturation of H to create an adequate driving force for the formation of dislocation loops. The α' can hence dissolve larger amounts of H (2-3 at.%) than pure α-Ti. However, the upper limit of H solubility in α' martensite is constrained by its hexagonal lattice. Early literature reported that Al alloying could increase H solubility in α and suppress hydride precipitation in Ti-alloys [37–39]. Our results agree with higher levels of H detected in Al-alloyed α than in unalloyed α, thereby confirming that Mo-supersaturation and Al-addition make hydride formation more difficult. Without hydride formation in alloyed α, H is confined to β, which, combined with a possible higher susceptibility for FIB-damage, explains why large amounts of H (~17 at.%) are detected in the β phase in Ti6246 and Ti64.

# 5   Conclusion

To conclude, we demonstrated that significant quantities of H can be measured in a variety of model and commercial Ti-alloys, in particular in the vicinity of α-grain boundaries and α/β interfaces, with local concentrations exceeding 10 at.%, and often as high as 50±2 at.% within hydrides. Macroscopic investigations of extrinsic properties are insufficient and only local characterisation, at the near-atomic scale, enables the detailed understanding necessary to explain engineering behaviour. In particular, when consulted for estimating the likelihood of hydride formation, the Ti – H phase diagram can be misleading as it is the local and not the global chemical potential that impacts hydride formation. Hydrides form along interfaces and boundaries in unalloyed α phase more easily than in alloyed α by Al or Mo oversaturation. The implication of our findings is that hydrogen embrittlement in commercial Ti alloys (with Al addition) near room temperature is probably not a consequence of hydride formation but the concentration of H into the beta phase that can promote local decohesion or enable enhanced localized plasticity facilitating accelerated nucleation of cracks. The direct observation of H opens future routes for material design to overcome hydrogen embrittlement and to improve the design of more efficient hydrogen storage materials.

**Acknowledgments**

Uwe Tezins & Andreas Sturm for their support to the FIB & APT facilities at MPIE. YC is grateful to the China Scholarship Council (CSC) for funding of PhD scholarship. AB, MH, DP, DR and BG acknowledge the Deutsche Forschungsgemeinschaft (DFG) for partly funding this research through SFB 761 "steels ab initio". J. Gussone, J. Haubrich, G. Requena from the Institute of Materials Research, German Aerospace Center (DLR), Linder Höhe, Cologne are thanked for providing the Ti64 alloy prepared by selective laser melting. Funding for AA is from EPSRC grant EP/M506345/1, with material supplied by Rolls Royce. MPM and PAJB acknowledge the financial support of the EPSRC under grants EP/L014742/1 and EP/M022803/1.

Figure Captions:

*Figure 1: scanning electron micrograph showing in (a) and the corresponding EBSD map in (b) the α' martensite structure in the as-quenched Ti-4Mo after homogenization at 950°C for 24 hours. (c) APT H and Mo distributions and patterns formed on the detector during the analysis exhibiting the typical symmetries the HCP phase for the as-quenched Ti-4Mo and (d) composition profile through the lath boundary.*

*Figure 2: (a) APT H and Mo distributions map and (b) composition profile through the lath boundary for the Ti-4Mo sample after 15min at 600°C to initiate β-reversion.*

*Figure 3: (a) bright-field TEM image of an APT tip from the annealed Ti – 4 wt.% Mo alloy, (b) selected area diffraction pattern (SADP) indexed as $[\bar{1}11]β$ // [011]hydride. (c) Simulated diffraction patterns along the bcc-β $[\bar{1}11]$ zone axis parallel to the hydride [011] zone axis. (d) dark-field (DF)-TEM micrographs, where hydride is coloured in purple, β in yellow and α in cyan. (e) Correlative APT analysis and (f) 1D compositional profile along the yellow arrow in (e) where the errors, represented as line shading, corresponds to the 2σ counting errors in each bin of the profile.*

*Figure 4: atom probe crystallography analysis of a hydride precipitated at a α low-angle grain boundary in Ti-2Fe: (a) patterns formed on the detector during the analysis exhibiting the typical symmetries from the local crystalline phase highlighted by the superimposed stereograms; (b) APT reconstruction and (c) spatial distribution maps revealing the presence of atomic planes in the tomographic reconstruction near the α-Ti / hydride interface shown in (d); (e) model of the faceted α/hydride interface.*

*Figure 5: (a) APT H and Ti distributions map and composition profile through the boundary of hydride and α matrix for CP Ti. (b) bright-field micrograph of a TEM lamellar of CP Ti, prepared by FIB, the inset is the selected area diffraction pattern (SADP) indexed as $[01\bar{1}0]α$ // [011]hydride, indices in yellow for hydride and in cyan for α matrix. (c) dark-field (DF)-TEM micrographs taken from $(11\bar{1})$ hydride marked in yellow in the diffraction pattern.*

*Figure 6: APT analyses of (a) Ti6246 and (b) Ti64 showing the H distribution and a composition profile across the α/β interface.*

*Figure 7: relative composition of $H_1^+$, $H_2^+$, $Ti^{2+}$ and $Ti^{3+}$ measured across the α/β interface in Ti6246. The blue dotted line is the ratio of $Ti^{3+}$ over all the other charge states of Ti.*

*Figure 8: H composition in APT datasets depending on the phases detected across all alloys analysed in our study represented by different symbols. H level in unalloyed α, i.e. in CP Ti and two phase binary model alloys, is ~1 at.%. More H is detected in alloyed α (by Al in Ti6246 and Ti64, and by Mo in quenched Ti-4Mo with martensite). β phases contain much higher amount of H than α phases in all alloys here. Hydride forms along α/β interface in binary alloys or α grain boundaries in CP Ti, while no hydride observed in commercial Ti alloys, Ti6246 and Ti64.*

Figure 1

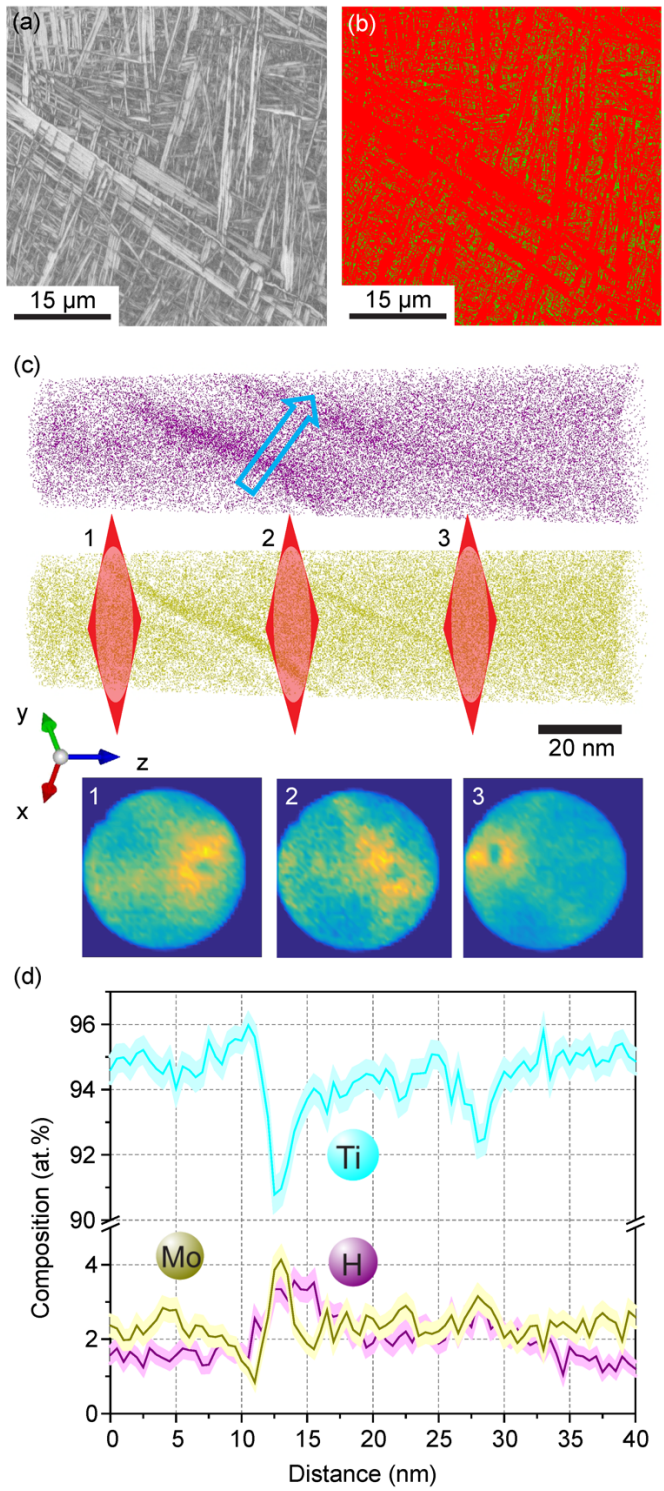

Figure 2

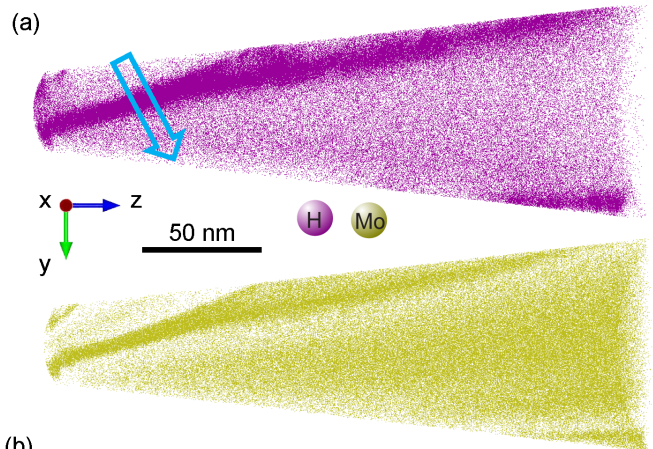

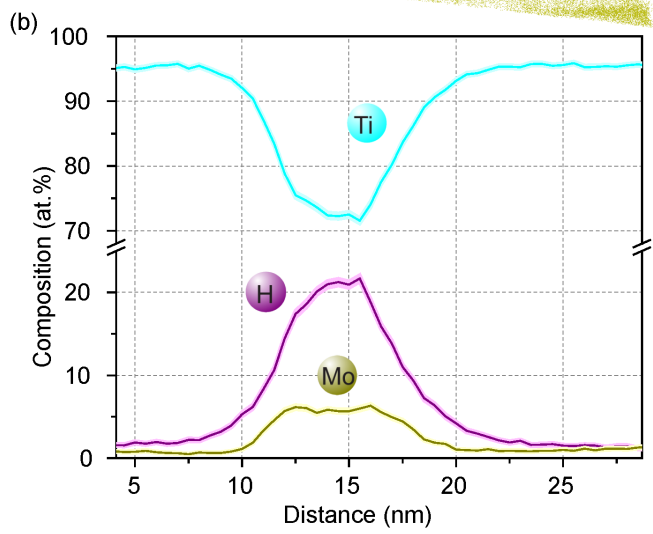

Figure 3

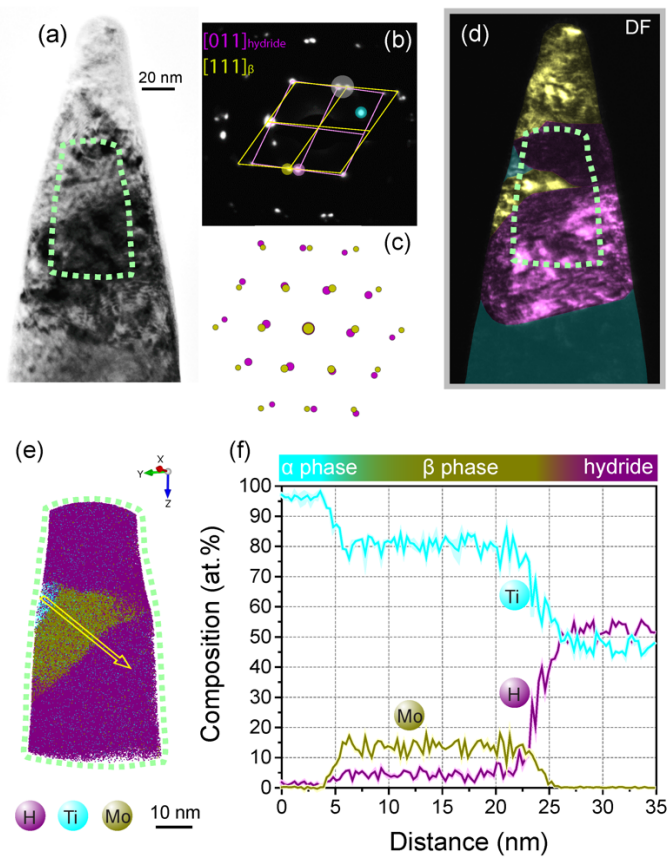

Figure 4

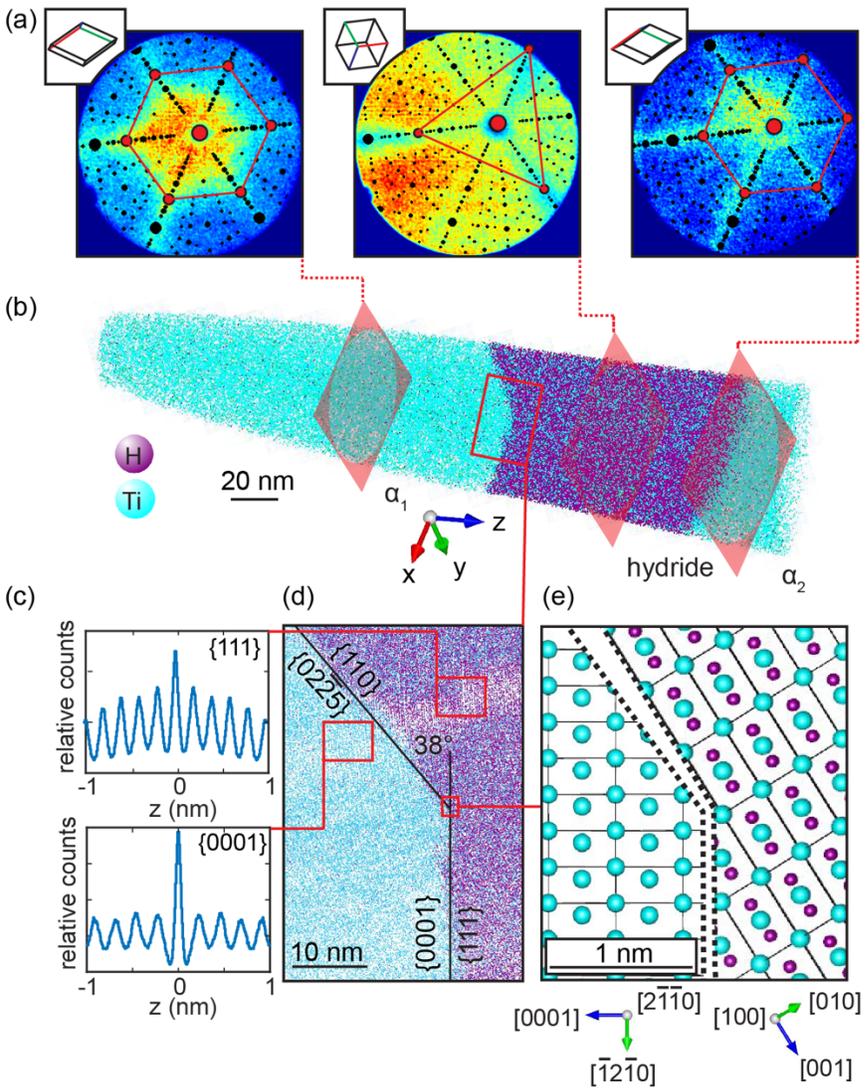

Figure 5

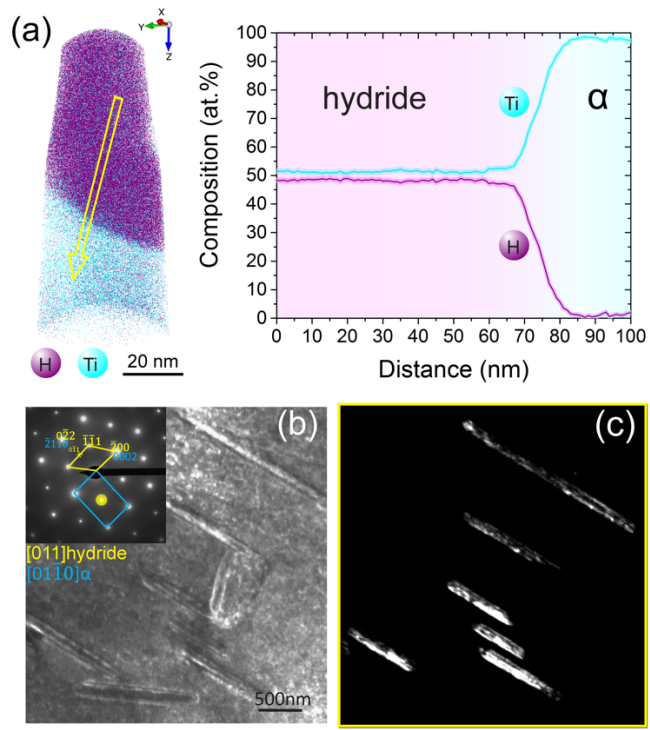

Figure 6

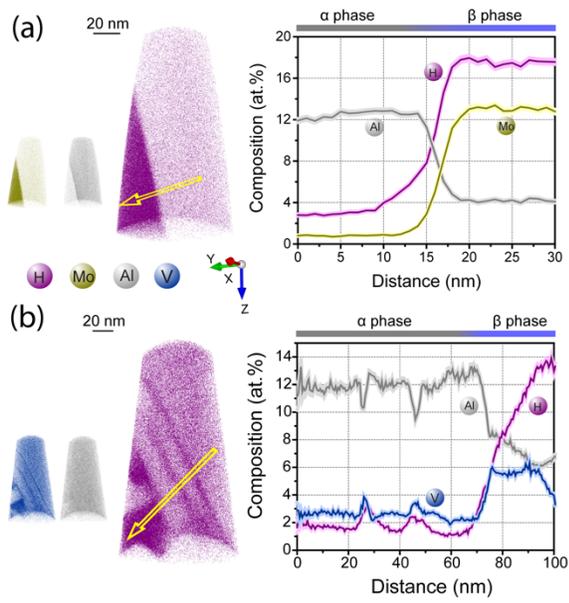

Figure 7

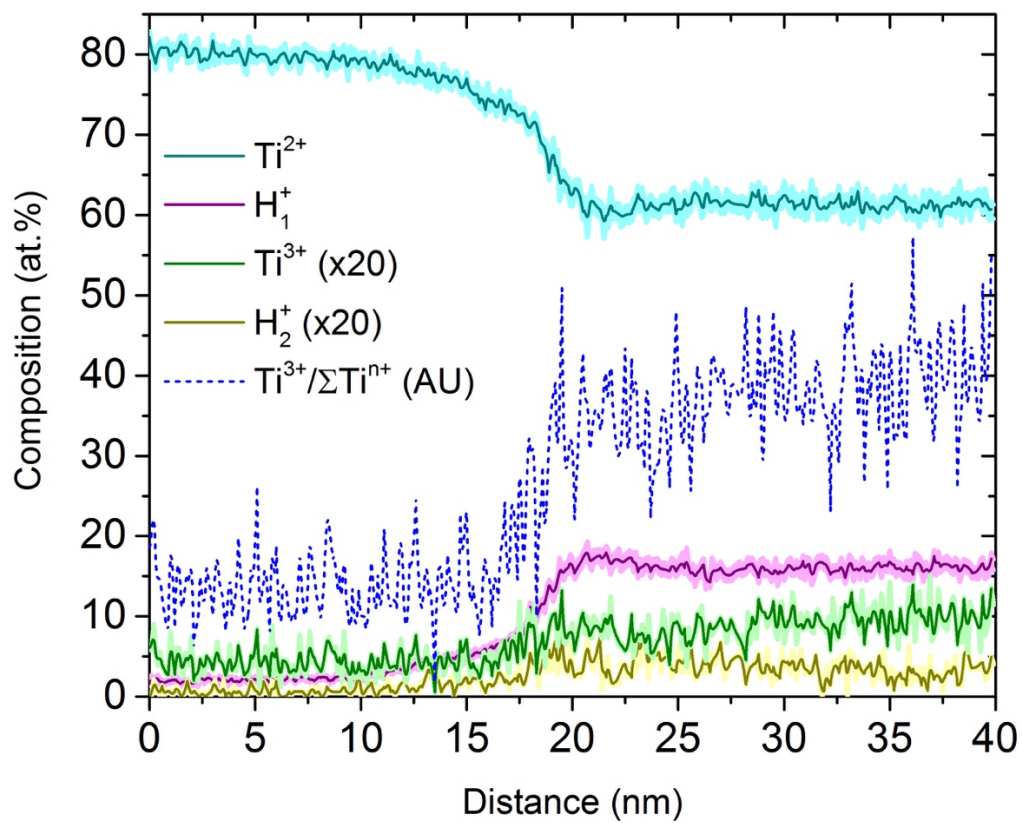

Figure 8

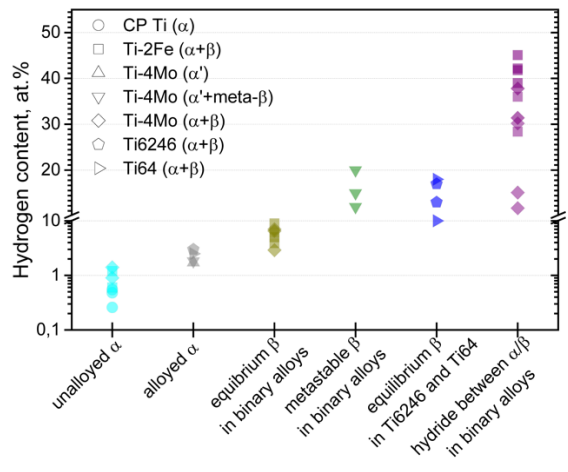